\documentclass{mystyle}

\usepackage{graphicx}  
\title{Blazars distance indications from {\it Fermi} and TeV data}
\author{E.Prandini\from{ins:x}\thanks{e-mail: prandini@pd.infn.it}\ETC,
G.~Bonnoli\from{ins:y}, L.~Maraschi\from{ins:z}, M.~Mariotti\from{ins:x}\\
        \atque F.~Tavecchio\from{ins:y}}
\instlist{\inst{ins:x} Dipartimento di Fisica \& INFN Padova, via Marzolo 8, 35134, Padova
          \inst{ins:y} Osservatorio Astronomico di Merate, via E. Bianchi 46, 23807, Merate (Lc)
          \inst{ins:z} Osservatorio Astronomico di Brera, via Brera 28, 20121, Milano}
\PACSes{
\PACSit{95.55.Ka}{X and $\gamma$-ray telescopes and instrumentation}\PACSit{98.54.Cm}{Active and peculiar galaxies and related systems}\PACSit{98.70.Vc}{Background radiations}
}
\begin{document}

\maketitle

\begin{abstract}
A new method to constrain the distance of blazars with unknown redshift using combined observations in the GeV and TeV regimes will be presented. The underlying assumption is that the Very High Energy (VHE) spectrum corrected for the absorption of TeV photons by the Extragalactic Background Light (EBL) via photon-photon interaction should still be softer than the extrapolation of the gamma-ray spectrum observed by Fermi/LAT. Starting from the observed spectral data at VHE, the EBL-corrected spectra are derived as a function of the redshift z and fitted with power laws. Comparing the redshift dependent VHE slopes with the power law fits to the LAT data an upper limit to the source redshift can be derived.

 The method is applied to all TeV blazars detected by LAT with known distance and an empirical law describing the relation between the upper limits and the true redshifts is derived. This law can be used to estimate the distance of unknown redshift blazars: as an example, the distance of PKS 1424+240 is inferred.
\end{abstract}

\section{Introduction}
\begin{center}
  \begin{table}
    {\small
      \centering
      \begin{tabular}{llcccccc}
        \hline
        Source Name     & $z$[real] &{\it Fermi}/LAT slope  & TeV slope      & $z^*$   \\%^ & $z$[rec]  \\         
     %                   &           & slope           &slope      &          &  \\ 
        \hline
        Mkn 421          & 0.030 & 1.78 $\pm$ 0.03        & 2.3 $\pm$ 0.1  & 0.08 $\pm$ 0.02 \\
        Mkn 501          & 0.034 & 1.73 $\pm$ 0.06        & 2.3 $\pm$ 0.1  & 0.10 $\pm$ 0.02 \\
        1ES 2344$+$514   & 0.044 & 1.76 $\pm$ 0.27        & 2.9 $\pm$ 0.1  & 0.20 $\pm$ 0.06\\
        Mkn 180          & 0.045 & 1.91 $\pm$ 0.18        & 3.3 $\pm$ 0.7  & 0.20 $\pm$ 0.12 \\
        1ES 1959$+$650   & 0.047 & 1.99 $\pm$ 0.09        & 2.6 $\pm$ 0.2  & 0.09 $\pm$ 0.04 \\
        BL Lacertae      & 0.069 & 2.43 $\pm$ 0.10        & 3.6 $\pm$ 0.5  & 0.23 $\pm$ 0.12 \\
        PKS 2005$-$489   & 0.071 & 1.91 $\pm$ 0.09        & 3.2 $\pm$ 0.2  & 0.19 $\pm$ 0.04 \\
        W Comae          & 0.102 & 2.02 $\pm$ 0.06        & 3.7 $\pm$ 0.2  & 0.23 $\pm$ 0.05 \\
        PKS 2155$-$304   & 0.116 & 1.87 $\pm$ 0.03        & 3.4 $\pm$ 0.1  & 0.22 $\pm$ 0.01 \\
        1ES 0806$+$524   & 0.138 & 2.04 $\pm$ 0.14        & 3.6 $\pm$ 1.0  & 0.23 $\pm$ 0.15 \\
        1ES 1218$+$304   & 0.182 & 1.63 $\pm$ 0.12        & 3.1 $\pm$ 0.3  & 0.21 $\pm$ 0.08 \\
        1ES 1011$+$496   & 0.212 & 1.82 $\pm$ 0.05        & 4.0 $\pm$ 0.5  & 0.49 $\pm$ 0.12 \\
        S5 0716$+$714    & 0.310$^{a,b}$ & 2.16 $\pm$ 0.04 & 3.4 $\pm$ 0.5  & 0.21 $\pm$ 0.09 \\
        PG 1553+113      & 0.400$^c$ & 1.69 $\pm$ 0.04    & 4.1 $\pm$ 0.2  & 0.57 $\pm$ 0.05 \\
        3C66A            & 0.444$^a$ & 1.93 $\pm$ 0.04    & 4.1 $\pm$ 0.4  & 0.34 $\pm$ 0.05 \\
        3C279            & 0.536 & 2.34 $\pm$ 0.03        & 4.1 $\pm$ 0.7  & 0.75 $\pm$ 0.72\\ 
        \hline
      \end{tabular}
      \caption
      {TeV blazars used in this study. The sources used in this study 
        are listed in the first column, their redshift (second column), their 
        {\it Fermi}/LAT slope (third column), the VHE slope of the observed differential energy
        spectrum fit (fourth column) and the value $z^*$ (last column). 
        $^a$Uncertain; $^b$from \cite{nilsson08}; 
        $^c$from \cite{danforth10}. Detailed references can be found in \cite{prandini10}.}\label{table_values}
    }
  \end{table}
\end{center}
The extragalactic TeV sky catalogue ($E>100$ GeV),
counts nowadays 45 objects\footnote{For an
updated list see: http://www.mppmu.mpg.de/$\sim$rwagner/sources/}. 
Many of these sources have been recently detected also at
GeV energies by the {\it Fermi} satellite~\cite{abdo09}, allowing 
for the first time a quasi-continuous coverage of the spectral shape 
of extragalactic VHE emitters over more than 4 decades of energy. 
The large majority of extragalactic TeV emitting objects
are blazars, radio-loud active
galactic nuclei with a relativistic jet closely oriented toward 
the Earth, as described in \cite{urry}.
Here, we discuss a method, recently published in \cite{prandini10}, 
to derive an upper limit on the redshift of 
a blazar, based on the comparison between the spectral index at GeV energies
as measured by LAT (unaffected by the cosmological absorption up to redshifts
far beyond those of interest here) and the TeV spectrum corrected for the 
absorption. Starting from the derived limits, we find a 
simple law relating these values to real redshift, which can 
be used to guess the distance of unknown redshift blazars.
We assume a cosmological scenario 
with $h=0.72$, $\Omega_M=0.3$ and $\Omega_\Lambda=0.7$.
\section{Results}
The photon flux emitted by a blazar in both GeV and TeV regimes
can be well approximated by power laws, of the form 
$dN/dE = f_0 (E/E_0)^{-\Gamma}$, where $\Gamma$ is the power-law index. 
At VHE, the photons of the spectrum interact with the extragalactic
background light (EBL), via electron-positron pair creation.
Quantitatively, the effect % of the interaction
is an exponential attenuation of the flux 
by a factor $\tau (E,z)$, where $\tau$
is the optical depth, a function of 
both photon energy and source redshift. Thus, the observed
differential energy spectrum from a blazar, $F_{\rm obs}$, 
is related to the emitted one, $F_{\rm em}$,  according 
to $F_{\rm obs}(E)=e^{-\tau(E)} F_{\rm em}(E)$.

In order to estimate a safe upper limit to the source distance,
we can reasonably assume that the intrinsic spectrum at TeV 
energies cannot be harder than that in the adjacent GeV band. 
Indeed, from the brightest objects studied at both GeV and TeV 
energies it appears that the SED is continuous, with a broad peak not requiring 
additional spectral components \cite{aharonian09}. 
Hence, a natural assumption is to require that the
slope measured in the GeV energy range is a 
limit value for the power-law index
of the de-absorbed TeV spectrum. 

For the study, we consider the blazar sample listed in table~\ref{table_values}
and containing all the extragalactic 
TeV emitters located at redshift larger than $z=0.01$, 
detected by LAT after taking 5.5 months of data \cite{abdo09}.
In order to estimate the redshift
$z^*$ for which the TeV spectral slope equals to
the GeV one, the measured spectral points of each source 
have been corrected for the corresponding 
absorption factor \cite{franceschini08}, starting from redshift $z=0.01$,
and the resulting spectrum fitted with a power law. 
The procedure, applied in fine steps of 
redshift, is iterated until the slope of the 
de-absorbed spectrum equals to the one 
measured by LAT. The corresponding redshift, $z^*$, reported in
table~\ref{table_values},
is the limit value on the source distance.

\begin{figure}
    \centering  
    \includegraphics[width=4.in]{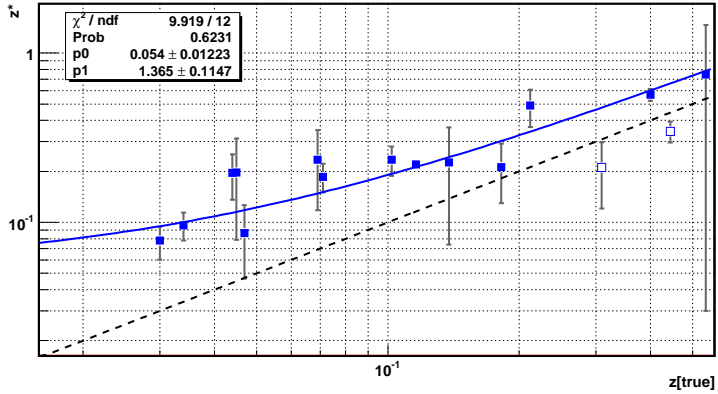}
    \caption{$z^*$ versus true redshift
      derived with the procedure described in the text.
      The open points are the two uncertain redshift
      sources, namely 3C~66A and S5~0716+714, not used in 
      the fit calculation (continuous line). The dashed line is the bisector.}
    \label{correlationplot}
  \end{figure}

Among the 16 sources considered in this study,
14 blazars have well-known redshift and are used to test
the method, while the remaining  
two blazars (3C~66A and S5~0716+714) have uncertain redshift, 
and are considered separately. 
The errors on $z^*$ are estimated taking into account both 
errors on the TeV and LAT slopes.
Fig.~\ref{correlationplot} shows the comparison between 
the known redshift, $x$-axis, and $z^*$.
All the $z^*$ lie above the bisector (dashed line) 
meaning that their values are larger than those of the the real redshift 
$z[true]$. This 
is expected since we are not considering the presence of the intrinsic break in
the blazar spectra, and {\it confirms that the method can be used to
set safe upper limits on blazars distance.} The only exceptions are the two sources 
with uncertain distance, S~0716+714 and 3C~66A (open circles).
This could be either due to some intrinsic properties of the sources or 
to a wrong estimate of their distances. 
In the latter case, our method would constrain, at two sigma level, 
the redshift of S5~0716+714 below $0.39$ and that of 3C~66A 
below $0.44$.

In \cite{stecker10}, a linear expression 
for the steepening of the observed TeV slope due to EBL absorption is derived.
Since in our procedure $z^*$ is related to this steepening,
it is natural to assume that also $z^*$ and $z$[true] 
are related by a linear function, of the form $z^*=A+Bz$[true].
The meaning of the coefficients is rather transparent: 
basically $A$ is a measure of the intrinsic 
spectral break of the sources, while, following \cite{stecker10},
$B$ is a measure (increasing values
for decreasing EBL level) of the optical depth of the EBL model used. 
We interpolate with this linear function 
the data with well-known distance of figure~\ref{correlationplot}.
The linear fit (continuous line) has a
probability of $62\%$.
Once derived this empirical relation, one can use it to
{\it determine the redshift} of sources with uncertain distance. 
For S5~0716+714 the reconstructed redshift 
is $z[rec]$~= 0.11~$\pm$~0.05, 
while that of 3C~66A is $z[rec]$~= 0.21~$\pm$~0.05. The error quoted is
estimated in \cite{prandini10}.
\section{The redshift of PKS~1424+240}
\begin{figure}
     \centering
     \includegraphics[width=3.5in]{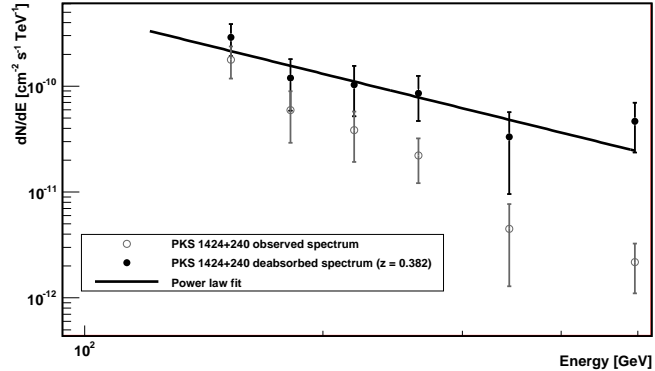}
     \caption{Measured (open points) and deabsorbed (filled points)
       spectrum of PKS~1424+240 at redshift $z$~=~0.382.}
     \label{1424_plot}
   \end{figure}

As a final example of application, we use our procedure 
on PKS~1424+240, a blazar of unknown redshift
recently observed in the VHE regime by Veritas \cite{acciari10}. 
The slope spectrum measured by {\it Fermi}/LAT  
between $0.2$ and $300$ GeV is 
$1.85\pm0.05$. The corresponding $z^*$
redshift at which the de-absorbed TeV spectrum slope
becomes equal to it, is $0.382\pm0.105$, figure~\ref{1424_plot},
using the EBL model \cite{franceschini08}.
This result is in agreement with the value of $0.5\pm0.1$, 
reported in \cite{acciari10},
calculated applying the same procedure 
but only simultaneous {\it Fermi} data.
Our estimate on the most probable distance for PKS~1424+240,
obtained by inverting the $z^*$ formula,
is $z$[rec]$=0.24\pm0.05$,  where, as before, the error quoted is
estimated in \cite{prandini10}.
\acknowledgments
GB, LM and FT acknowledge a 2007 Prin-MIUR grant for financial support.

\end{document}